\documentclass{PoS}

\usepackage{lineno}
\usepackage{amsmath}

\title{Charm Physics and CKM}

\ShortTitle{Charm Physics}

\author{\speaker{Carla G\"obel}\\
        Pontif\'\i cia Universidade Cat\'olica do Rio de Janeiro (PUC-Rio)\\
        E-mail: \email{carla.gobel@puc-rio.br}}

\abstract{An overview of the most important progresses in charm physics since the last CKM Workshop (2014) is presented. Due emphasis is given to the
experimental measurements directly related to the CKM matrix.}

\FullConference{9th International Workshop on the CKM Unitarity Triangle\\
		28  November - 3 December 2016\\
		Tata Institute for Fundamental Research (TIFR), Mumbai, India}

\begin{document}
%\linenumbers
\section{Introduction}
Studies of charmed mesons and baryons play an important role in the understanding of weak and strong interactions. With masses around 2 GeV, open-charm hadrons pose a theoretical challenge for the description of hadronic transitions. From the experimental side, it offers richness from the multiplicity of hadronic, semi-leptonic and leptonic channels to study.

In what concerns the Cabibbo-Kobayashi-Maskawa (CKM) matrix more directly, the charm sector is unique. Charm transitions at tree level are regulated by the 2$\times$2 Cabibbo sub-space of the CKM matrix, which is almost unitary and real (deviations of ${\cal O}(\lambda^4)$). Nevertheless, loop (``penguin") processes involving flavour-changing-neutral-current (FCNC) transitions make it possible the phenomena of $D^0$--$\overline{D}^0$ mixing and CP violation. Although the former is already firmly established, there still is no experimental evidence for the latter.  Thus so far this scenario points to consistency with  Standard Model (SM) predictions, where CP violation asymmetries are expected to be tiny \cite{Bianco:2003vb, Grossman:2006jg}. New Physics (NP) contributions could considerably enhance the effect. Although theoretical expectations are tough to estimate, a safe statement  is that CP asymmetries approaching the percent level or higher are hard to be explained by the SM.

Currently, Charm Physics is benefiting from large experimental samples, providing precise measurements through a few key modes. Yet, the effort on the search for CP violation and tests of the CKM unitarity in the charm sector have to be based on a large number of decay modes. 

We present herein an overview of the latest achievements in Charm Physics. The main focus is on the experimental results since the CKM Workshop in 2014.

\section{Direct CP Violation}

Direct CP violation occurs when the amplitude for the decay of a particle into a given final state  differs from that of its charge conjugate state, 

\begin{equation}
{\cal A}(D\to f) \neq \overline{\cal A} (\overline{D}\to \bar f)~.
\end{equation}

Within the SM, it is necessary to have at least two interfering amplitudes, with both weak and strong phase differences, for CP violation to be observed. Direct CP violation in charm can  arise in singly-Cabibbo-suppressed (SCS) decays for which both tree and penguin amplitudes contribute. 
%These transitions at quark level are shown in Fig.~\ref{SCSDiagrams}. 
The most important penguin amplitude involves a $c\to b\to u$ transition which naively has a CKM suppression of $\lambda^4$ ($\lambda \approx 0.225$) with respect to the tree amplitude. Estimates for  direct CP asymmetries with charm are thus very small,  roughly  ${\cal O}(10^{-3})$ or less \cite{Bianco:2003vb, Grossman:2006jg}.

\subsection{Time-integrated asymmetries}

In two-body decays and for phase-space integrated measurements in multi-body decays, direct CP violation is searched for through the asymmetry observable 

\begin{equation}
A_{\rm CP} = \frac{\Gamma(D\to f) - \Gamma(\overline{D} \to \bar f)}{\Gamma(D\to f) + \Gamma(\overline{D} \to \bar f)}~, 
\end{equation}
where $\Gamma(D \,(\overline{D})\to f)$ represents the decay width of the $D$ ($\overline{D}$) meson to a final state $f$.

For the SCS decays\footnote{Charge conjugation is implied unless stated explicitly.} $D^0\to h^-h^+$ ($h = \pi, K$), the final state is a CP eigenstate and an interesting measurement is the asymmetry difference defined as

\begin{equation}
\Delta A_{\rm CP} = A_{\rm CP}(D^0\to K^-K^+) - A_{\rm CP} (D^0\to \pi^-\pi^+)~,
\end{equation}
which is mostly a measure of direct CP violation \cite{Gersabeck:2011xj}.

This measurement has generated, a few years ago, a lot of excitement  \cite{Aaij:2011in, Collaboration:2012qw, Ko:2012px} with the belief to be the first evidence for  CP violation in charm. But LHCb has since then brought news, with results coming from independent samples based on two $D^0/\overline{D}^0$ tagging techniques. The first relies on the decay chain   $D^{*+}\to D^0(hh)\pi^+$ and it's  called here as {\it $D^*$}-tagging. The second is based on b-hadron decays, $B\to D^0 X\mu$ where the charge of the muon determines the $D$ flavour, the so-called $\mu$-tagging.  With the full run I data (3 fb$^{-1}$ of pp collisions at 7 and 8 TeV),  and using the $\mu$-tagged sample, LHCb has found \cite{Aaij:2014gsa} $\Delta A_{\rm CP} = (0.14\pm 0.16\pm 0.08)\%$ \footnote{Here, unless stated otherwise, when two uncertainties are quoted, the first one is the statistical uncertainty and the second is the systematic uncertainty. When a single uncertainty is quoted, it is the combination of both.}. The awaited update with the $D^*$-tagged sample has been released in 2016. The result is \cite{Aaij:2016cfh}
$ \Delta A_{\rm CP} = (-0.10\pm 0.08\pm 0.03)\%.$

Even more recently, LHCb has also provided values for the individual CP asymmetries \cite{Aaij:2016dfb}, $A_{\rm CP}(KK)=(0.04\pm 0.12\pm 0.10)\%$ and 
$ A_{\rm CP} (\pi\pi)= (0.07\pm 0.14\pm 0.11)\%$.
The sensitivities in these channels have reached $10^{-3}$ with  the results showing no sign of CP violation. 

Searches in other channels, however, still suffer from larger uncertainties. Among them, an interesting and important channel with recent results is $D^0\to K^0_S K^0_S$. For this decay, CP violation is expected to be enhanced even through CKM effects only. Indeed, a theoretical upper limit reaches 1.1\% \cite{Nierste:2015zra}.  LHCb, with full run I data sample, measured recently $A_{\rm CP} (D^0\to K^0_S K^0_S) = (2.9\pm 5.2\pm 2.2)\%$ \cite{DKsKs_LHCb}. Currently, the most precise measurement comes from Belle \cite{Abdesselam:2016gqq}: $A_{\rm CP} (D^0\to K^0_S K^0_S) = (-0.2\pm 1.53\pm 0.17)\%$. We can see that the precision is still higher than the upper theoretical limit but getting close. Good expectations lie on this channel from run II LHCb data.
 
Belle also brings results on radiative decays of the type $D^0 \to V\gamma$ ($V$ being a vector resonance).  These decays are sensitive to NP effects through  chromomagnetic dipole operators. Belle observes the decay $D^0\to \rho^0\gamma$ for the first time and measures \cite{Abdesselam:2016yvr} $A_{\rm CP}(D^0\to \phi\gamma) = (-9.4\pm 6.6\pm 0.1)\%$, $A_{\rm CP}(D^0\to K^{*0}\gamma) = (-0.3\pm 2.0\pm 0.0)\%$ and $A_{\rm CP}(D^0\to \rho^0\gamma) = (5.6\pm 15.2\pm 0.6)\%$.

 A few results come also from charged $D$'s.  LHCb has recently studied the decays $D^+_{(s)}\to \eta'\pi^+$, where $\eta'\to \pi^-\pi^+\gamma$,  and found $A_{\rm CP}$ consistent with zero for both $D^+$ and $D_s^+$ \cite{Aaij:2017eux}, with a precision (combining statistical and systematic uncertainties) of about 0.9\% and 0.5\%, respectively. BESIII has analysed $D^+$ with final states $K^0_{L,S}K^+(\pi^0)$ \cite{ZhaoCharm2016} and $K^0_L e^+\nu_e$ \cite{Ablikim:2015qgt} also finding consistency with the non-CP violation hypothesis, with uncertainties ${\cal O}(\%)$. 
 
 Thus, apart from the $D^0\to h^+h^-$ final states, just a few channels so far have reached sensibility well below the  1\% level for integrated $A_{CP}$ asymmetries. There is yet plenty of room for NP searches.

\subsection{Multi-body decays}
 
 In multi-body decays, CP violation can be studied throughout the decay phase-space. In a 3-body spinless decay, the phase space is bi-dimensional and usually represented by the Dalitz plot, where the independent variables are 2-body squared-mass combinations. For higher number of final-state particles and/or baryon decays, the  phase space dimension is at least five.  The advantage of performing a search for CP violation within the phase space is that typically one should expect local asymmetries to be  larger than the phase-space integrated ones. This is a sensible argument considering that CP observables are a consequence of the dynamics through possible interfering amplitudes, and this could be enhanced by resonances giving clear signatures in the phase space. 
 
The search for local asymmetries can be done through model-dependent (amplitude analysis) and model-independent techniques. The latter relies on a direct  comparison of the phase-space distribution of events for particle and anti-particle decays. One such a way is the binned method already used in a few analysis over the past few years \cite{Bediaga:2009tr}. A relatively novel method is an unbinned approach, the {\it energy test} \cite{Williams:2011cd}, used for the first time by LHCb in $D^0\to \pi^-\pi^+\pi^0$ \cite{Aaij:2014afa}. In both methods, a p-value of consistency with the CP-conservation hypothesis is obtained, which can be converted to a significance for CP violation. As such, these are discovery methods; in case of observation of CP violation,  there will be  the need of a full amplitude analysis for a quantitative evaluation of the CP asymmetries with their corresponding sources. For 4-body  or baryon decays, one can also construct triple products of the type $C_T = \vec p_3\cdot(\vec p_1\times \vec p_2)$ (with $\vec p_i$, $ i =1,2,3$ denoting the momenta of particles in the final state) to probe for CP violation (see, for instance, \cite{Aaij:2014qwa}).

Two recent results are based on full amplitude analyses to search for CP violation. LHCb has studied the decays $D^0\to K^0_S K^\pm\pi^\mp$ \cite{Aaij:2015lsa} while a CLEO data-legacy analysis was performed on $D^0 \to \pi^-\pi^+\pi^-\pi^+$ decays \cite{dArgent:2016rbp}. By analyzing separately $D^0$ and $\overline{D}^0$ Dalitz plots, the extraction of magnitudes, phases and fit fractions are done, and in particular $A_{\rm CP}$ is obtained for each resonant channel. LHCb results for  $D^0\to K^0_S K^\pm\pi^\mp$ presented uncertainties on CP asymmetries ranging from 0.4\% to a few \%, depending on the intermediate state. $D^0\to 4\pi$ analysis presented asymmetries for the various intermediate states with uncertainties not below 3\%. 

With a much higher sample than that from CLEO, LHCb has applied the energy-test method for $D^0\to \pi^-\pi^+\pi^-\pi^+$ decays \cite{Aaij:2016nki}. From simulation, sensitivity for CP violation effects (at least a $3\sigma$ inconsistency with the CP-conservation hypothesis) are found to be typically 4--5\% in amplitudes or 3--4$^\circ$ in phases for the main intermediate states. Using triple products, it is possible also to separate CP-even and -odd samples to test for CP violation. Results show p-values for CP-conservation hypothesis  of $(4.3\pm 0.6)\%$ and $(0.6\pm 0.2)\%$ for CP-even and -odd studies, respectively. The CP-odd result is only marginally consistent with CP conservation.   

\section{Mixing and CP Violation}

As for the neutral $K$ and $B$ mesons, $D$ mesons also exhibit mixing, being the only up-quark type system where the effect is possible. The weak eigenstates are written as $|D_{1,2}\rangle = p |D^0\rangle \pm q |\overline{D}^0\rangle$, with well defined masses and widths $m_{1,2}$ and $\Gamma_{1,2}$, respectively. The observables related to mixing are $x = \frac{2(m_2-m_1)}{\Gamma_1+\Gamma_2}$ and $y = \frac{\Gamma_2-\Gamma_1}{\Gamma_1+\Gamma_2}$. Indirect CP violation occurs if $|p/q|\neq 1$ or $\phi = \arg (p/q)\neq 0$. A way to independently measure $p, q$ and $\phi$, as well as $x$ and $y$,  is through  a time-dependent Dalitz plot analysis, for instance through the decay $D^0 \to K^0_S\pi^+\pi^-$. 

One can also define CP observables related to mixing that can be measured from $D\to hh$ decays
\begin{eqnarray}
%A_\Gamma = \frac{\hat\Gamma(D^0\to f) - \hat\Gamma(\overline{D}^0\to f)}{\hat\Gamma(D^0\to f) + \hat\Gamma(\overline{D}^0\to f)} = \frac{1}{2} \left[\left(\left|\frac{q}{p}\right| - \left|\frac{p}{q}\right|\right) y\cos\phi - \left(\left|\frac{q}{p}\right| + \left|\frac{p}{q}\right|\right)x\sin\phi\right]~, \nonumber \\
%y_{CP} = \frac{\hat\tau(D^0\to K^-\pi^+) }{\hat\tau(D^0\to K^-K^+) } -1 = \frac{1}{2} \left[\left(\left|\frac{q}{p}\right| + \left|\frac{p}{q}\right|\right) y\cos\phi - \left(\left|\frac{q}{p}\right| - \left|\frac{p}{q}\right|\right)x\sin\phi\right]
A_\Gamma &=& \frac{1}{2} \left[\left(\left|\frac{q}{p}\right| - \left|\frac{p}{q}\right|\right) y\cos\phi - \left(\left|\frac{q}{p}\right| + \left|\frac{p}{q}\right|\right)x\sin\phi\right]~, \nonumber \\
y_{CP} &=& \frac{1}{2} \left[\left(\left|\frac{q}{p}\right| + \left|\frac{p}{q}\right|\right) y\cos\phi - \left(\left|\frac{q}{p}\right| - \left|\frac{p}{q}\right|\right)x\sin\phi\right]~.
\end{eqnarray}
In the absence of CP violation, $y_{\rm CP} = y$ and $A_\Gamma = 0$. When measuring time-dependent asymmetries,  one has  $A_{\rm CP} (t) \approx a_{\rm CP}^{\rm dir} + a_{\rm CP}^{\rm ind}\frac{\langle t\!\rangle}{\tau}$, where $\langle t\!\rangle$ is the measured average decay time, $\tau$ is the decay lifetime, and with $a_{\rm CP}^{\rm ind} = -A_\Gamma$ being the contribution from indirect CP violation and $a_{\rm CP}^{\rm dir}$ the term from direct CP violation \cite{Gersabeck:2011xj}.

Evidence for $D^0-\overline{D}^0$ mixing appeared for the first time about a decade ago \cite{Aubert:2007wf, Staric:2007dt,Aaltonen:2007ac}, with the first observation from a single experiment coming in 2012 \cite{Aaij:2012nva}. The effect is  now  firmly established. The measurements has usually relied on the study of $D^0\to K^+\pi^-$ and $D^0\to K^-\pi^+$ -- ``wrong-sign'' (WS) and ``right-sign'' (RS) decays respectively. The WS decay can occur through a  doubly-Cabibbo-supressed (DCS) decay amplitude and via a $D^0-\overline{D}^0$  oscillation followed by a Cabibbo-favored (CF) decay, while the RS decay ocurs via a CF decay and via $\overline{D}^0-D^0$ oscillation followed by a DCS decay. Assuming $x, y\! \ll \!\!1$, the ratio between the  yields of the WS and the RS decays  varies in time as
\begin{equation}
R(t)^\pm = R_D^\pm + \sqrt{R_D^\pm} y'^\pm \left(\frac{t}{\tau}\right) + \frac{(x'^\pm)^2+(y'^\pm)^2}{4}\left(\frac{t}{\tau}\right)^2~,
\end{equation}
where $t/\tau$ is the decay time  measured in units of the average $D^ 0$ lifetime $\tau$.
The parameters $x', y'$ relate to $x, y$ by a rotation due to the the strong-phase difference between the DCS and the CF amplitudes. $R_D^ \pm$ are  the ratios between  the DCS and the CF decay rates, where the $\pm$ superscripts refer to the measured quantities for $D^0$ and $\overline{D}^0$, respectively, when the study is performed allowing for CP violation. 

Interesting news on mixing measurements came very recently. BaBar has performed the first measurements of $x$ and $y$ using a time-dependent amplitude analysis of the decay $D^0\to \pi^+\pi^-\pi^0$. Although, as expected, the precision is not competitive to that from $K^0_S\pi^+\pi^-$ decays \cite{delAmoSanchez:2010xz,Peng:2014oda} it is an excellent news to see more actors entering the game. The same goes with the results from LHCb in a model-independent analysis of $D^0\to K^0_S\pi^+\pi^-$ with just 1\,fb$^{-1}$: the precision is not great, but it serves mainly as a proof-of-principle on the alternative way-to-go in the future, since Dalitz modeling adds irreducible systematics. The method relies on input of the strong phase differences across the Dalitz plot from $D^0-\overline{D}^0$ quantum-correlated pairs in the $\psi(3770)$ production, as provided by CLEO \cite{Briere:2009aa}. On this subject, the very good news come from BESIII which has recently presented preliminary results on these strong phases \cite{Weidenkaff:2016mgr}, with precision much improved with respect to that from CLEO. 

On another front, LHCb has just released a new analysis on $D^0-\overline{D}^0$ mixing by including {\it doubled-tagged} events, {\it i.e.} from the decay chain $B \to \mu^-D^{*+}X$, $D^{*+}\to D^0(K\pi)\pi^+$ \cite{Aaij:2016roz}. These  correspond to a very clean, independent sample from the previous 3\,fb$^{-1}$ analysis based on prompt sample \cite{Aaij:2013wda}, providing further sensitivity on low decay proper-time. Overall the reported uncertainties on the fit parameters (for no CPV model as well as CP-allowed models) are improved by 10--20\%. 

Last but not least, for the first time mixing has been observed in a 4-body decay! LHCb used $D^0\to K^+\pi^-\pi^+\pi^-$ to obtain the WS to RS ratio as a function of decay time and the non-mixing hypothesis was excluded with $8.2\sigma$. The ratio amplitude, $r_D^{K3\pi}$, and the product of the coherence factor by the rotated mixing parameter, $R_D^{K3\pi}\cdot y'_{K3\pi} $, are measured, which represent important input for the measurement of the CP-violating phase $\gamma$ in $B\to DK$ decays \cite{Harnew:2013wea,Harnew:2014zla}.

\begin{figure}
\begin{center}
\includegraphics[width=0.45\textwidth]{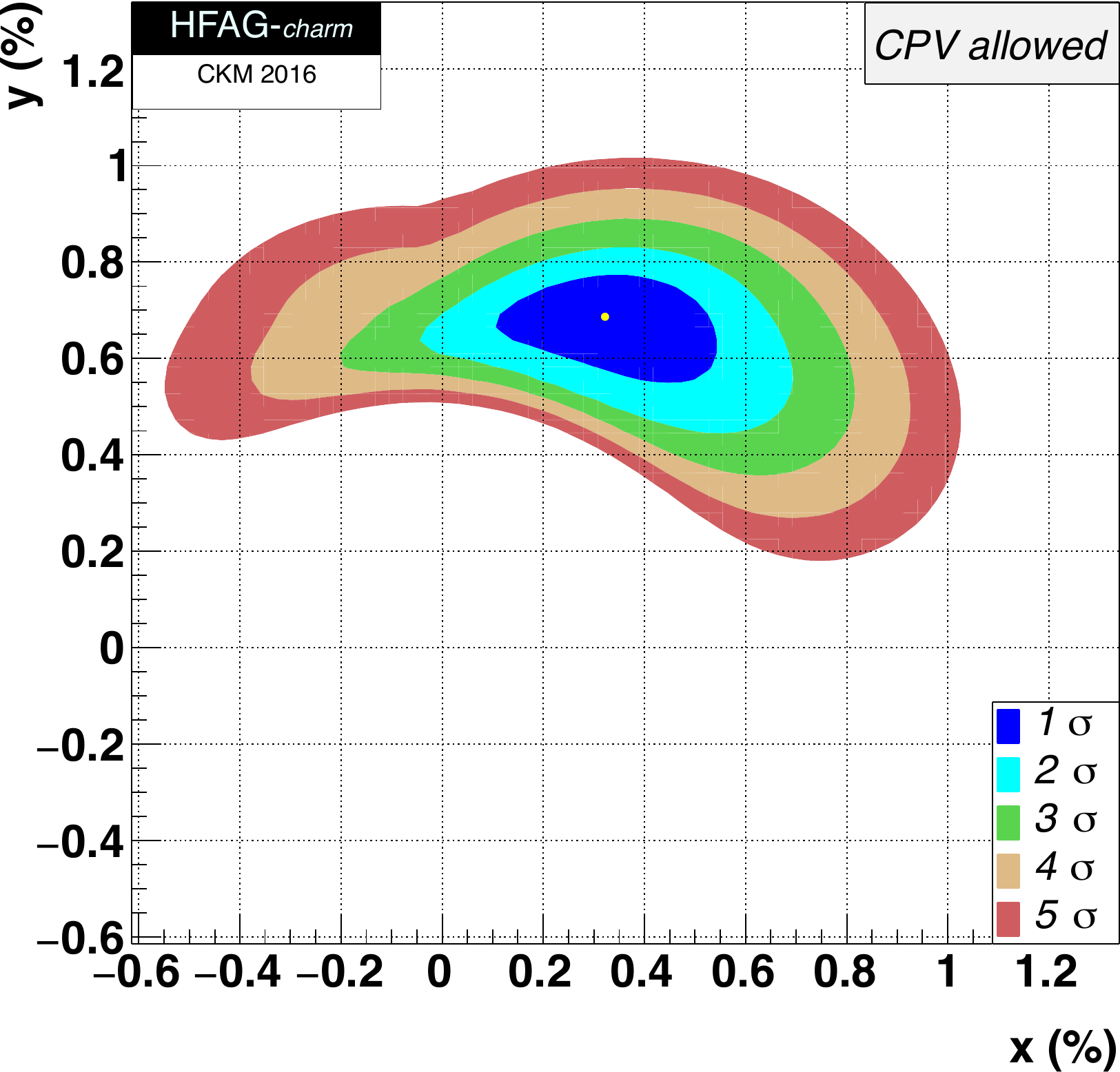}
\includegraphics[width=0.45\textwidth]{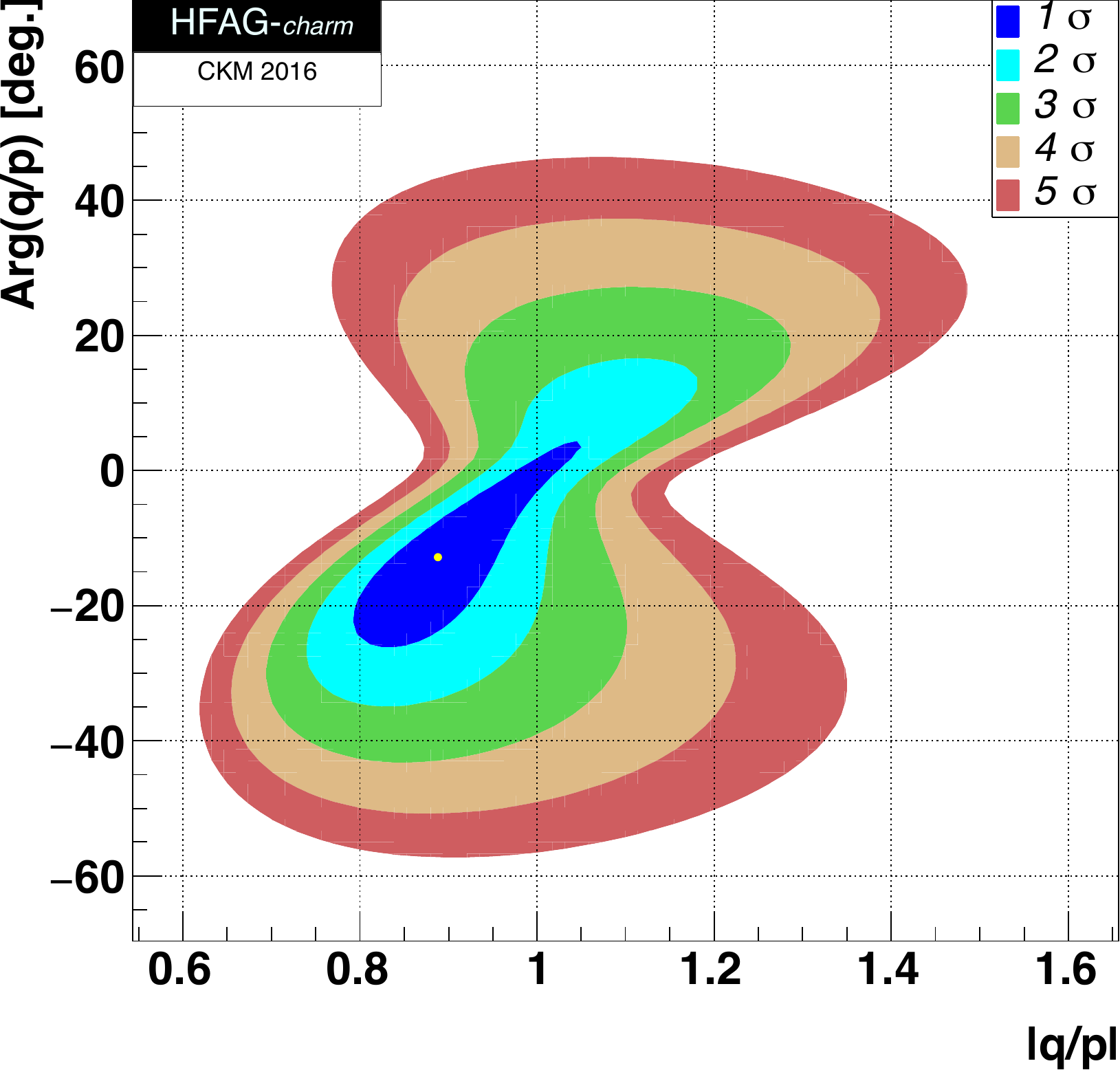}

\caption{Global fit plots for $D^0-\overline{D}^0$ mixing observables from HFAG \cite{Amhis:2016xyh}.}
\label{hfag-plots}
\end{center}
\end{figure}

The status of mixing  in the charm sector is well summarized by the global fit performed by HFAG \cite{Amhis:2016xyh}, updated for CKM 2016, and reproduced here in Figs.~\ref{hfag-plots}.

Moving to  indirect CP violation, BESIII has results on $y_{\rm CP}$ using CP-tagging techniques in $D\overline{D}$ quantum-correlated pairs: a $D^0$ is CP-tagged, via selected CP eigenstates, while the other decay semi-leptonically $D\to K\ell\nu$ \cite{Ablikim:2015hih}. Assuming no direct CP violation, BESIII finds $y_{\rm CP} = (-2.0\pm 1.3\pm 0.7)\%$.  

But again the most significant contributions on CP violation in mixing come from $D\to h^+ h'^-$ decays. Belle has its final dataset results released \cite{Staric:2015sta}. From a simultaneous fit to $D^0\to K^-\pi^+$, $D^0\to K^-K^+$ and $D^0 \to \pi^-\pi^+$, they measure $y_{\rm CP} = (1.11\pm 0.22\pm 0.09)\%$ and $A_{\Gamma} = (-0.03\pm 0.20\pm 0.07)\%$. These results show consistency with CP conservation, with $y_{\rm CP}$ statistically comparable to $y$ and $A_\Gamma$ consistent with zero. 

The most precise measurement of $A_\Gamma$ comes from LHCb  with the full run I dataset with $D^*$-tagged sample \cite{Aaij:2017idz}. The analysis was performed with two methodologies: unbinned maximum likelihood fits to obtain the effective lifetimes, using per-event acceptances; and a binned method, where the time-dependent CP asymmetry is studied in proper decay time bins. The first method is applied to data collected in 8 TeV collisions (2 fb$^{-1}$ collected in 2012) and then combined with the previous from 2011 data at 7 TeV (1 fb$^{-1}$) \cite{Aaij:2013ria}. The second method is applied to the whole 3 fb$^{-1}$ of data,  and results in a precision slightly better. For both methods, a pseudo $A_\Gamma(K\pi)$ is obtained as a control observable, and was found to be compatible with zero with a precision of $10^{-4}$.  The final results, based on the binned method,  are $A_\Gamma (KK) = (-0.30\pm 0.31\pm 0.14)\times 10^{-3}$ and $A_\Gamma (KK) = (0.46\pm 0.58\pm 0.16)\times 10^{-3}$. Combining the results for both channels, LHCb obtains 
$$A_\Gamma = (-0.12\pm 0.21\pm 0.10)\times 10 ^{-3}~, $$
which corresponds to the most precise CP violation measurement  so far. 

\begin{figure}
\begin{center}
\label{HFAG-yCPAGamma}
\includegraphics[width=0.45\textwidth]{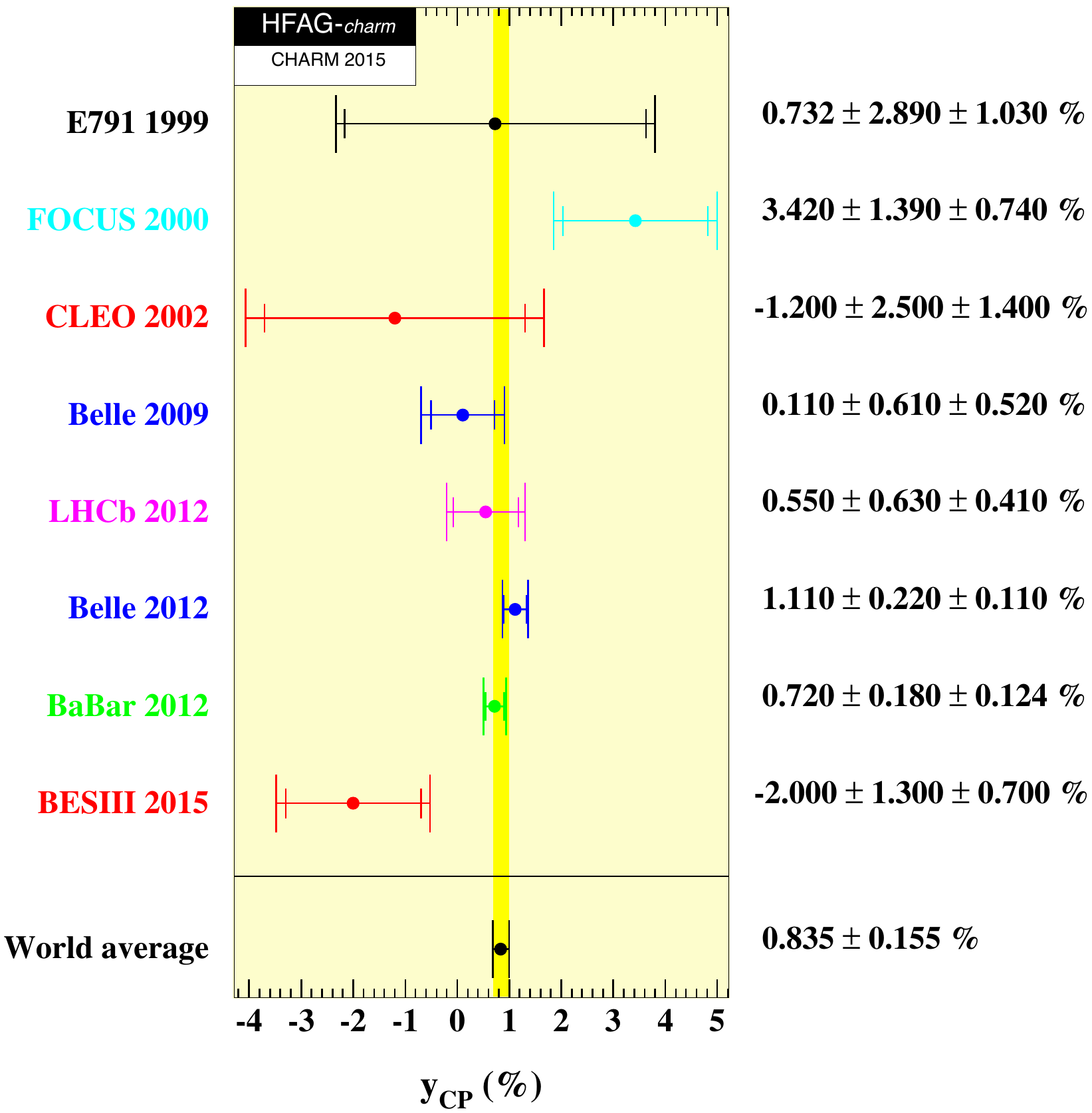}\hspace*{0.1\textwidth}
\includegraphics[width=0.45\textwidth]{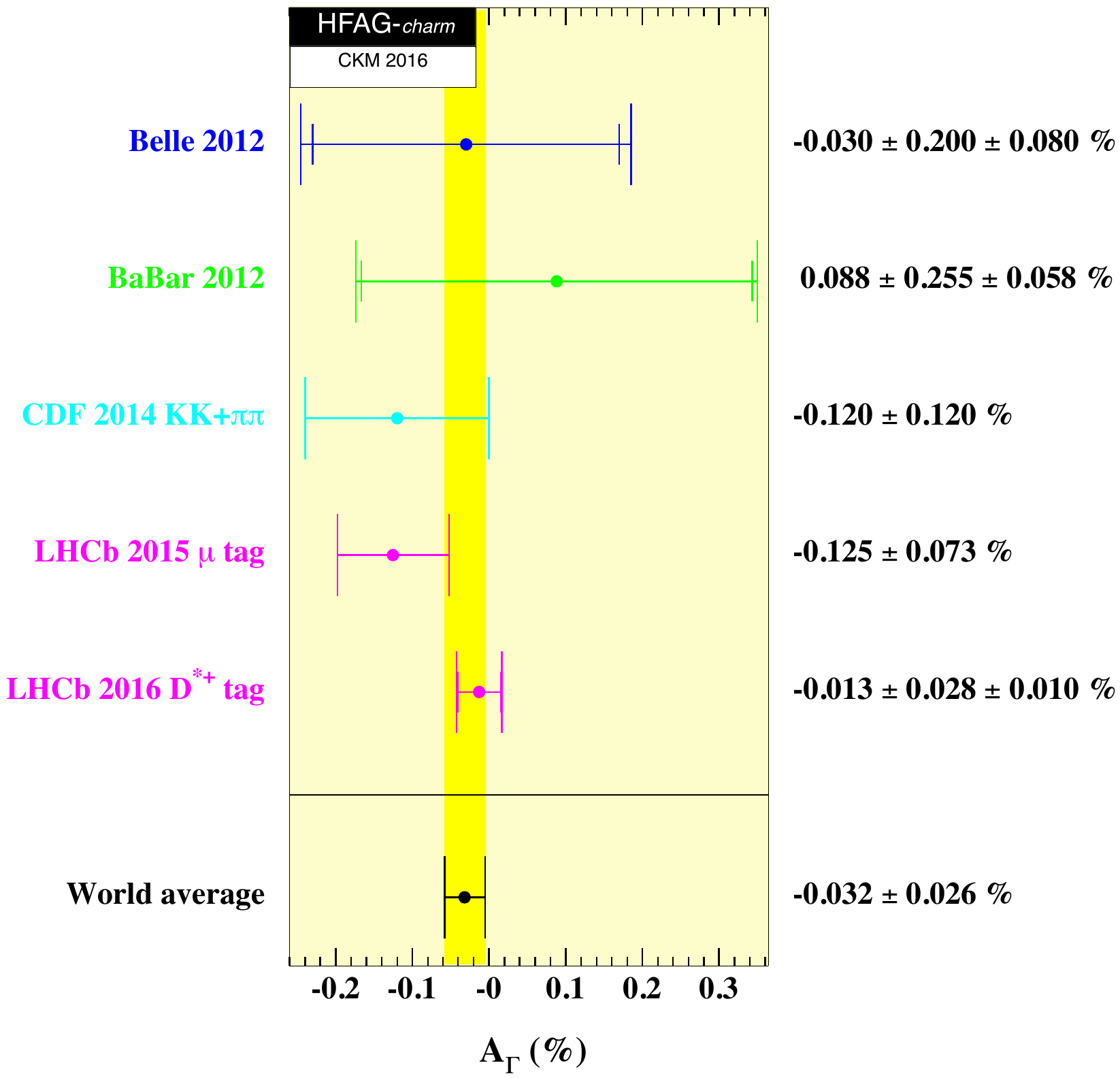}
\caption{$y_{\rm CP}$ and $A_\Gamma$ world averages from HFAG \cite{Amhis:2016xyh} .}
\end{center}
\end{figure}

\begin{figure}
\begin{center}
\label{HFAG-aDir-aInd}
\includegraphics[width=0.6\textwidth]{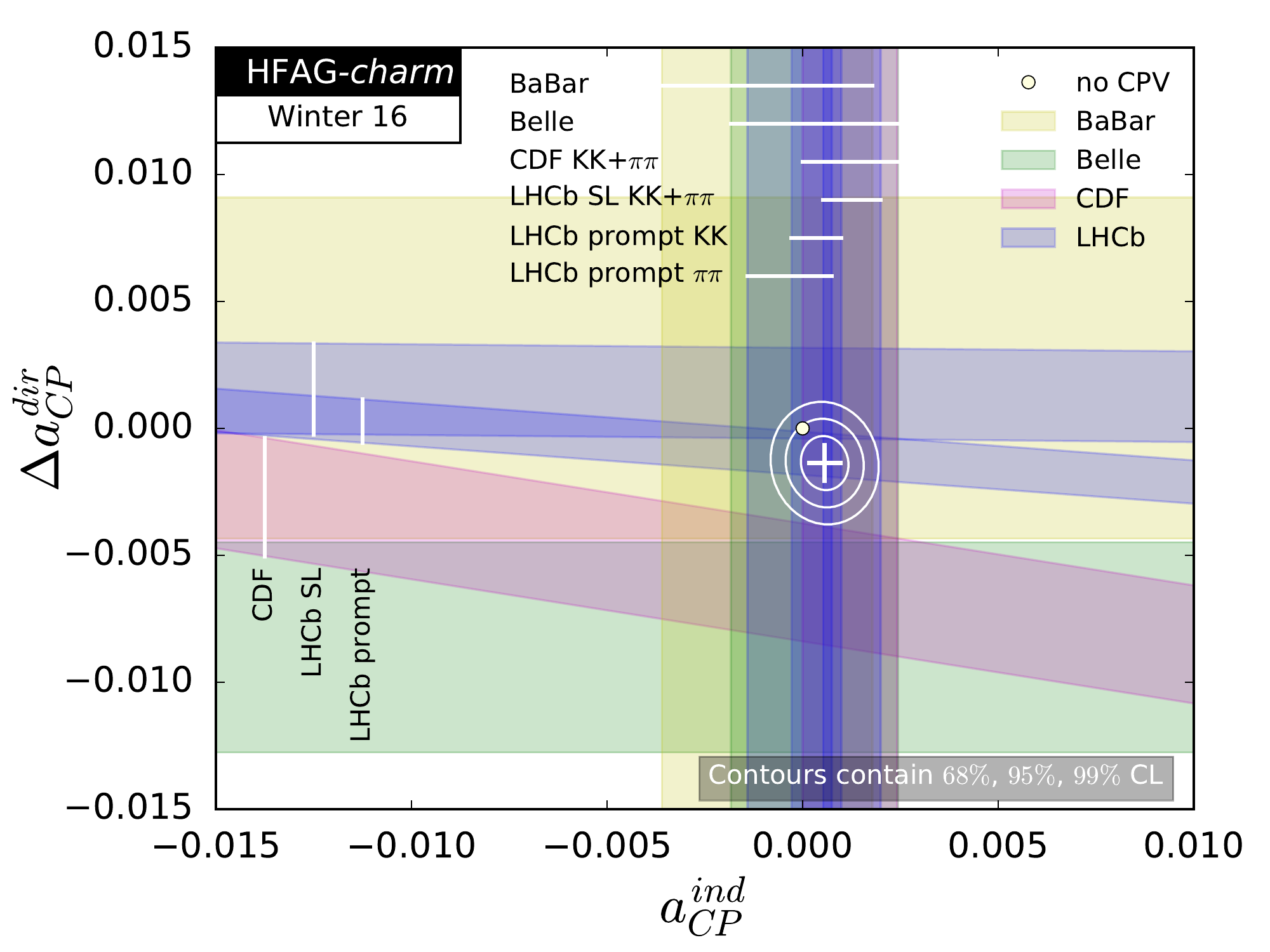}
\caption{Direct $\Delta a^{\it dir}_{\rm CP}$ vs. indirect $a^{\it ind}_{\rm CP}$ combination plot from HFAG \cite{Amhis:2016xyh}.}
\end{center}
\end{figure}

The world average results for $y_{\rm CP}$ and $A_\Gamma$, as obtained by HFAG \cite{Amhis:2016xyh}, are shown in Fig.~\ref{HFAG-yCPAGamma}. With the information coming from $\Delta A_{\rm CP}$, the interplay between direct and indirect CP violation from $D\to hh$ decays  can be seen in Fig.~\ref{HFAG-aDir-aInd} (also from HFAG).

\section{Advances in $\mathbf{V_{cd}}$ and $\mathbf{V_{cs}}$}

The Cabibbo sector of the CKM matrix is driven by sizeable elements of ${\cal O}(\lambda)$ and ${\cal O}(1)$. Nonetheless, precision measurements are fundamental under the scope of testing the SM predictions, in particular CKM unitarity. Leptonic and semi-leptonic charm decays enable the measurement of $|V_{cq}|$,  $q=(d,s)$. For leptonic decays, the observable is the branching fraction given by
\begin{equation}
{\cal B}(D^+_q \to \ell^+\nu_\ell) = \frac{G_F^2}{8\pi} \tau_{D_q} f^2_{D_q} |V_{cq}|^2 m_{D_q} m_\ell^2 \left( 1 - \frac{m_\ell^2}{m^2_{D_q}}\right)~,
\end{equation} 
where $G_F$ is the Fermi constant, $m_\ell$ is the fermion mass ($\ell = \mu, e$), and $m_{D_q}$, $\tau_{D_q}$ and $f_{D_q}$ are  the $D_q$ mass, lifetime and decay constant, respectively. 

In semi-leptonic decays of the type $D\to P\ell^+\nu_\ell$, with $P$ being a pseudo-scalar meson, one measures the differential decay rate given by
\begin{equation}
\frac{d\Gamma(D\to P\ell^+\nu_\ell))}{dq^2d\cos\theta_\ell} = \frac{G_F^2}{32\pi^3}p^{*3} |V_{cq}|^2 |f_+(q^2)|^2\sin\theta_\ell^2 ~,
\end{equation}
where $q$ is the 4-momentum of the lepton pair, $p^*$ is the  magnitude of the momentum of the $P$ meson in the $D$ rest frame and $\theta_\ell$ is the angle of the lepton direction in the $\ell\nu_\ell$ rest frame with respect to the direction of the $P$ meson in the $D$ rest frame. 

From the experimental side, one measures the product of the magnitude of $V_{cq}$ and $f_{D_q}$ or $f_+(q^2 = 0)$. Theoretical inputs, coming from Lattice QCD, are thus critical for the extraction of $V_{cq}$. A fairly complete compilation of the advances in Lattice QCD at low energy was released  by FLAG \cite{Aoki:2016frl} very recently.

Recent experimental results on leptonic decays comes from BESIII for both $D^+$ \cite{Ablikim:2013uvu}  and $D_s^+$ \cite{Ablikim:2016duz} mesons. This latest publication reports the total branching fractions for $D_s^+\to \mu^+\nu_\mu$ and $D_s^+ \to \tau^+ \nu_\tau$. BESIII brings also very precise measurements in semi-leptonic decays, 
$D^0\to \pi^-e^+\nu_e$ and $D^0\to K^-e^+ \nu_e$ \cite{Ablikim:2015qgt} , and  $D^+ \to  K_L^0 e^+\nu_e$ \cite{Ablikim:2015ixa}.  BaBar \cite{Lees:2014ihu} has also provided results for the $D^0\to \pi^-e^+\nu_e$ channel.

The most recent update from HFAG \cite{Amhis:2016xyh}, including all available data including the results above, gives the average values of $|V_{cd}|= 0.216\pm 0.005$ and $|V_{cs}| = 0.997 \pm 0.017$. Indirect measurements, from a global fit assuming CKM unitarity, give $|V_{cd}|= 0.22529^{+0.00041}_{-0.00032}$  and $|V_{cs}| = 0.973394^{+0.000074}_{-0.000096}$ which are consistent, within uncertainties, with the respective direct results quoted.

\section{Charm Baryons, briefly}

Although most of the results on charm physics typically comes from $D$ mesons, charm baryons can also provide important information on weak and strong dynamics. Recently, there has been very good news in this area. The first comes from Belle, which has published the first observation of the DCS decay $\Lambda_c^+ \to p\pi^+K^-$ \cite{Yang:2015ytm}. Then BESIII has brought many  important measurements. The branching ratios of the SCS decays $\Lambda_c^+ \to p \pi^+\pi^-$ and $\Lambda_c^+ \to p K^+K^-$ relative to the CF decay have been measured \cite{Ablikim:2016tze}. Twelve absolute branching fractions of 2-, 3- and 4-body CF hadronic decays of $\Lambda_c^+$, including $\Lambda_c^+\to pK^+\pi^-$, used for normalization, are obtained \cite{Ablikim:2015flg}, as well as the branching fraction of the semi-leptonic decay $\Lambda_c^+ \to \Lambda\mu^+\nu_\mu$ \cite{Ablikim:2016vqd}. Also, the decay $\Lambda^+_c\to nK^0_S\pi^+$ is observed for the first time  \cite{Ablikim:2016mcr}.

It would be natural and very welcome to expect charm baryons to play a more important role in the coming years. In particular, within the CKM context, it is worth mentioning the importance of the search for CP violation in $\Lambda_c^+$ SCS decays,  such as $\Lambda_c^+ \to ph^+h^-$, and the role of $\Lambda_c^+$ decays in providing important inputs for $\Lambda_b$ physics (including $V_{ub}$). 

\section{Conclusions}

Since CKM 2014, we've seen a surge in the number of very interesting results in Charm Physics. With many analyses with full run I dataset, LHCb has provided important results in mixing and CP violation. In  particular, the results on $\Delta A_{\rm CP}$ and $A_\Gamma$, on direct and indirect CP violation, respectively, have brought the sensitivities at the level of SM expectations. Mixing is now not only firmly established but also observed in a 4-body decay.  BaBar and Belle are also still providing nice results on these subjects. BESIII is appearing in very important fronts, making full use of the advantages of $D - \overline{D}$ quantum correlations in $\psi(3770)$.

Although for some observables the experimental sensitivities are already at the level of SM expectations, a comprehensive search for CP violation is the way to go: look everywhere. Indeed, there are still plenty of very interesting channels for which current sensitivities are well above the percent level. 

The next few years, with new results coming from BESIII, LHCb and soon Belle II, will hopefully  witness very exciting news from Charm Physics.

\bibliographystyle{JHEP}
\providecommand{\href}[2]{#2}\begingroup\raggedright\endgroup

%\end{thebibliography}

\end{document}